\documentclass[5p,twocolumn]{elsarticle}

\usepackage{graphicx}
\usepackage{dcolumn}
\usepackage{pifont}
\usepackage{bm}
\usepackage{multirow}
\usepackage{amsmath}
\usepackage{float}
\usepackage{txfonts}
\usepackage[version=3]{mhchem}

\usepackage[usenames,dvipsnames]{xcolor}
\definecolor{blue}{RGB}{0,0,225}
\definecolor{cream}{RGB}{222,217,201}
\definecolor{red}{RGB}{225,0,0}

\journal{Elsevier}

\begin{document}
\title{Defect physics in \ce{Yb^{3+}}-doped \ce{CaF2} from first-principles calculation}

\author[kimuniv-m]{Yun-Hyok Kye}
\author[kimuniv-m]{Chol-Jun Yu\corref{cor}}
\cortext[cor]{Corresponding author}
\ead{cj.yu@ryongnamsan.edu.kp}
\author[kimuniv-m]{Un-Gi Jong}
\author[kimuniv-m]{Chol-Nam Sin}
\author[jilin]{Weiping Qin}

\address[kimuniv-m]{Chair of Computational Materials Design, Faculty of Materials Science, Kim Il Sung University, Ryongnam-Dong, Taesong District, Pyongyang, Democratic People's Republic of Korea}
\address[jilin]{State Key Laboratory on Integrated Optoelectronics, College of Electronic Science and Engineering, Jilin University, Changchun, Jilin 130012, China}

\begin{abstract}
Calcium fluoride has been widely used for light up-/down-conversion luminescence by accommodating lanthanide ions as sensitizers or activators.
Especially, Yb-doped \ce{CaF2} exhibits unique defect physics, causing various effects on the luminescence.
This makes it vital for high efficiency of devices to control the defect-clustering, but theoretically principal guidelines for this are rarely provided.
Here we perform the first-principles study on defect physics in Yb-doped \ce{CaF2} to reveal the thermodynamic transition levels and formation energies of possible defects.
We suggest that the fluorine rich growth condition can play a key role in enhancing the luminescence efficiency by facilitating the Yb-clustering and suppressing the defect quenchers in bulk.
Detailed energetics of defect aggregation not only well explains the experimentally favored Yb-clustering but also presents $n$- or $p$-type doping method for the cluster control.
\end{abstract}

\maketitle

Recently, lanthanide (Ln) doped nanophosphors have attracted a great interest due to their appealing applications on photodynamic therapy, sensing, solar cell devices and so on~\cite{Huang_apl, Boyer_dds, Ai_natcomm, Wong_nanoscale2019, Huang_csr}.
Among various Ln hosting materials, fluorides have been paid much attention for their low phonon energies, excellent photochemical stability and low cytotoxicity~\cite{Zhou_jmcc, Li_rsc, Prorok_nanoscale}.
In particular, calcium fluoride (\ce{CaF2}) is one of the most intensively studied host for phosphors with interesting Ln-dopant clustering and up/down-conversion luminescent properties~\cite{Catlow_prb, Petit_jpc, Petit_prb, Lyberis_opt}.

Trivalent ytterbium ion (Yb$^{3+}$) is a typical Ln-dopant for \ce{CaF2}, used as a sensitizer for up-conversion (UC)~\cite{Qin_rsc1, Qin_pccp, Qin_lumin, Dong_nanoscale} while an activator for down-conversion (DC) processes~\cite{Molina_jap,Wijngaarden_prb}.
In the \ce{CaF2} host, Yb$^{3+}$ ions are prone to form clusters such as dimer, trimer, tetramer and hexamer, which enlarge the absorption and emission bands being beneficial to the efficient cooperative and non-cooperative sensitization~\cite{Ricaud_opt, Serrano_josab2012}.
Moreover, due to a single long-lived excited state of $^{2}$F$_{5/2}$, the Yb$^{3+}$ ions hardly suffer from the cross relaxation~\cite{Chen_acc, Chen_adv, Serrano_josab2011}, although other Ln-clusters often show optical losses due to the cross-relaxation.
In order to reveal the mechanism of such Ln clustering in \ce{CaF2}, many theoretical works have been reported~\cite{Catlow_prb, Catlow_jssc}, but yet a precise quantitative calculation of defect energetics at the level of quantum theory is not provided.

In this study, we investigate the defect physics of Yb-doped \ce{CaF2} by using density functional theory (DFT) calculations.
We make modeling of point and pair defects using the supercells, and calculate their formation energies and thermodynamic transition levels under the different growth conditions.
The ground and optically excited states of Yb$^{3+}$ ion are shown to be placed between the valence band maximum (VBM) and the conduction band minimum (CBM) of the host \ce{CaF2}, giving rise to the deep trap states for charge carriers being adverse to the efficient luminescence.
Through the analysis of defect energetics, we find that the F-rich (or Ca-poor) growth condition is favorable for suppressing the defects with deep trap states and for promoting the advantageous clustering of Yb-dopants.
Based on the results, we also suggest that $n$- or $p$-type doping can be used to manage the clustering of Ln-dopants.

\begin{figure}[!th]
\centering
\includegraphics[clip=true,scale=0.18]{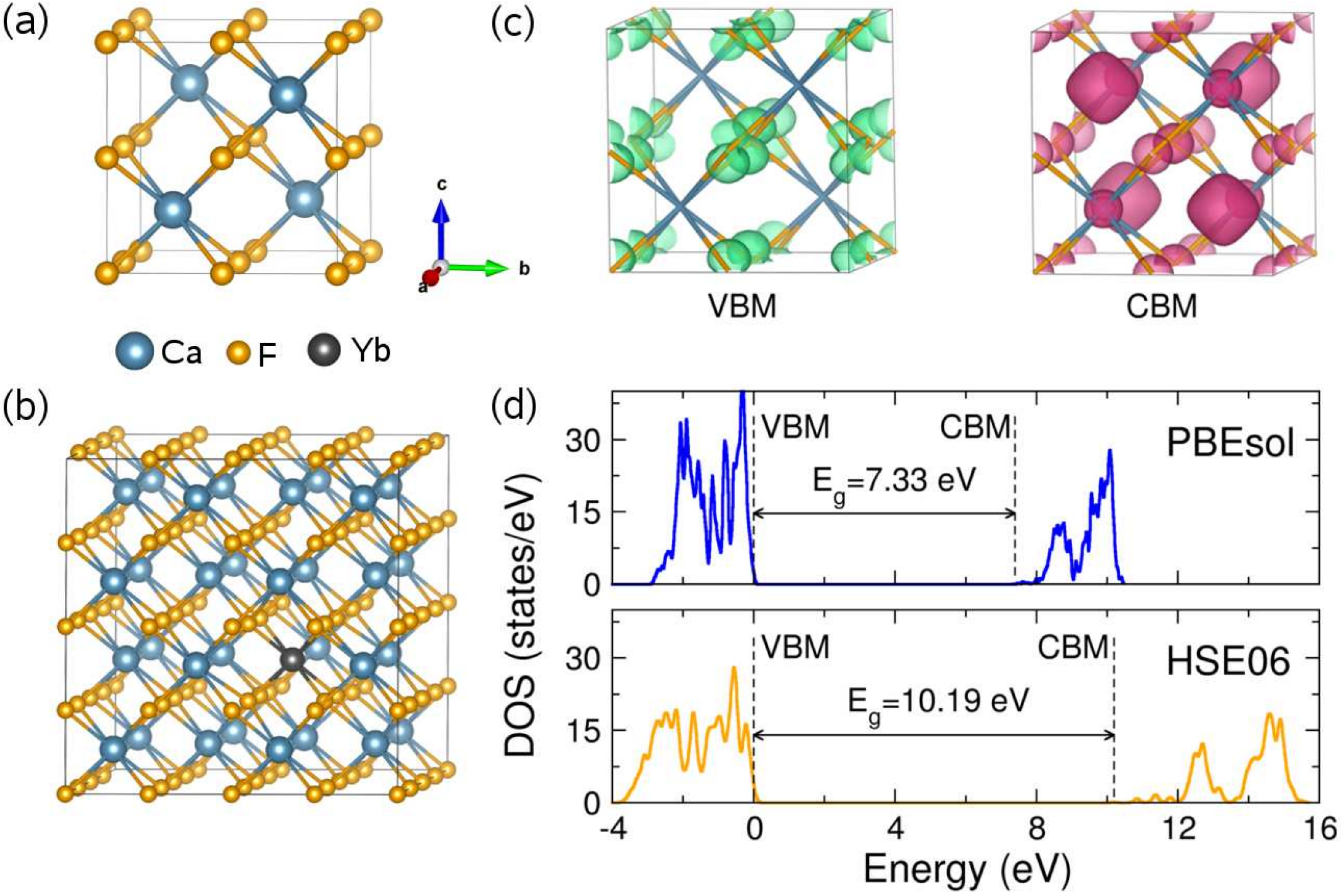}
\caption{\label{fig1}(a) Crystal structure of \ce{CaF2} unit cell and (b) $2\times2\times2$ supercell containing antisite defect Yb$_\text{Ca}$. (c) Isosurface plot of charge densities corresponding to the valence band maximum (VBM) and the conduction band minimum (CBM). (d) Density of states (DOS) in \ce{CaF2} unit cell with different band gaps $E_\text{g}$ from PBEsol and HSE06 functionals.}
\end{figure}

All the calculations were performed by using the projector augmented wave method as implemented in the Quantum ESPRESSO code~\cite{QE}.
The configurations of valence electrons are Ca-$3s^{2}3p^{6}4s^{2}$, F-$2s^{2}2p^{5}$, and Yb-$5s^{2}5p^{6}4f^{14}6s^{2}$, emphasizing the direct use of $f$ electrons in Yb element, and the scalar relativistic corrections were considered for the ionic cores.
The cutoff energies were set to be 40 and 400 Ry for wave function and electron density, respectively.
The special $k$-points were set to be $\Gamma$-centered $(8\times8\times8)$ and $(2\times2\times2)$ for \ce{CaF2} unit-cell (space group $Fm\overline{3}m$, 12 atoms, see Fig.~\ref{fig1}a) and Yb-doped \ce{CaF2} $2\times2\times2$ supercell (96 atoms, see Fig.~\ref{fig1}b).
Atoms were relaxed until the forces converged to $5\times10^{-5}$ Ry/Bohr with the PBEsol functional~\cite{PBEsol}.
After atomic relaxation, the DFT total energies were refined using the hybrid HSE06 functional~\cite{hse2} with 50\% nonlocal Hartree-Fock exchange addition, considering the underestimation of \ce{CaF2} band gap at the level of generalized gradient approximation (GGA) and the severe self-interaction error (SIE) of Yb $4f$ orbitals~\cite{Silva_prb, Janesko_pccp}.
In the calculation of defect formation energy, the finite size effects were corrected by using the calculated static dielectric constant of 8.0 and deep-lying Ca-3s level for the potential alignment~\cite{Walle, Kye_2019}.

%

The lattice constant of \ce{CaF2} unit-cell was determined to be 5.40 \AA~in good agreement with the experimental value of 5.45~\AA~\cite{Batchelder}.
As expected, the band gap was underestimated to be 7.33 eV with PBEsol-GGA, and widened to be 10.19 eV with hybrid HSE06, being closer to the experimental value of 11.80 eV~\cite{Rubloff_prb} (see Fig.~\ref{fig1}d).
With the optimized unit cell parameters, we built the 2$\times$2$\times$2 supercells, and made the point defects including interstitials (Ca$_\text{i}$, F$_\text{i}$, Yb$_\text{i}$), vacancies (V$_\text{Ca}$, V$_\text{F}$) and antisite (Yb$_\text{Ca}$) and pair defects (Yb$_\text{i}$-F$_\text{i}$, Yb$_\text{Ca}$-F$_\text{i}$, Yb$_\text{Ca}$-2F$_\text{i}$, 2(Yb$_\text{Ca}$-F$_\text{i}$)) in it (see Figs. S1 and S2).
Possible charge states were assigned to the defects.
Here, Yb$_\text{Ca}$-F$_\text{i}$ and 2(Yb$_\text{Ca}$-F$_\text{i}$) were devised to represent the experimentally observed Yb monomer and dimer in line with the Catlow's former theoretical work~\cite{Catlow_prb}.
Note that although the Yb$^{3+}$ dopants were thought to substitute the Ca$^{2+}$ ions, we also simulate the interstitial defect Yb$_\text{i}$, which is likely to form under the proper environmental condition.

\begin{figure*}[!th]
\centering
\includegraphics[clip=true,scale=0.55]{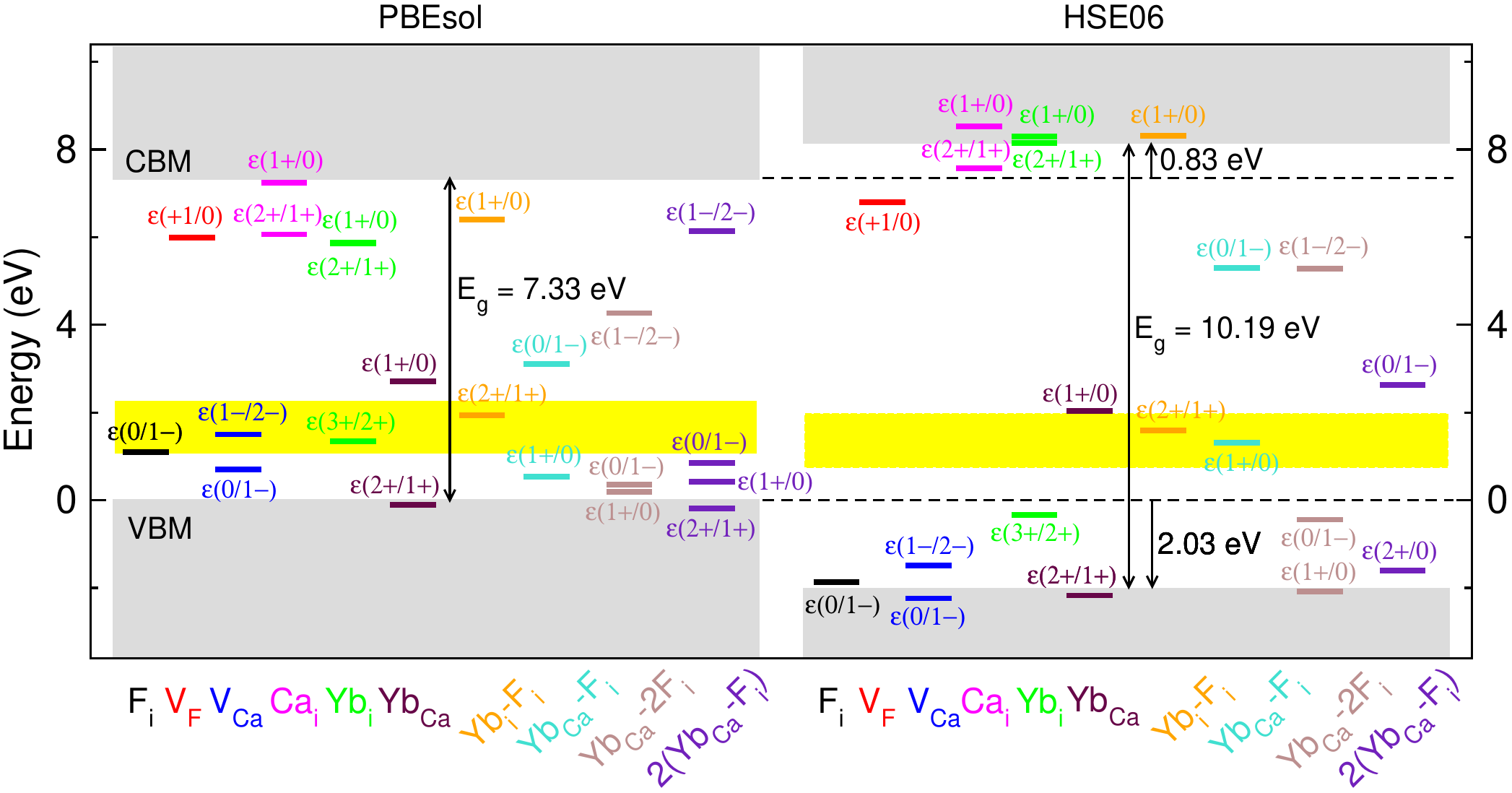}
\caption{\label{fig2}Thermodynamic transition levels of point and pair defects in Yb-doped \ce{CaF2} calculated with PBEsol (left) and HSE06 (right) functionals. Bands are aligned with respect to their electrostatic potentials. Yellow regions represent the Fermi energy values between the ground ($^{2}$F$_{7/2}$) and optically excited ($^{2}$F$_{5/2}$) levels of Yb$^{3+}$ ion.}
\end{figure*}

First, we evaluated the thermodynamic transition levels between the different charge states $q_1$ and $q_2$ of defect $D$ by applying the following equation~\cite{Walle, Agiorgousis_jacs},
\begin{equation}
\varepsilon(q_{1}/q_{2})=\frac{E[D^{q_{1}}]+E_\text{corr}[D^{q_{1}}]-E[D^{q_{2}}]-E_\text{corr}[D^{q_{2}}]}{q_{2}-q_{1}} \label{eq_1}
\end{equation}
where $E[D^q]$ is the DFT total energy of the supercell containing the defect with charge state $D^q$ and $E_\text{corr}[D^q]$ is the correction term for the finite size effect of supercell and defect-induced electrostatic potential shift~\cite{Walle}.
Figure~\ref{fig2} shows the thermodynamic transition levels for the point and pair defects calculated by both PBEsol and HSE06 functionals.
When using the hybrid HSE06 functional, VBM is shifted downward by 2.03 eV being similar to the former calculation of 1.80 eV with the hybrid PBE0~\cite{Ibraheem_eur}, while CBM is shifted upward by 0.83 eV.
We see that in the PBEsol calculation all the native point defects have deep transition levels in the region between VBM and CBM, whereas in HSE06 calculation most of the transition levels move towards the band edges becoming shallower, except the V$_\text{F}$ defect.

In order to determine which levels can trap excited 4$f$ electrons of Yb$^{3+}$ ions during the energy transferring process, we correlate the Fermi energy of defective supercells containing nominally charged Yb$^{3+}$ ion to the ground $^{2}$F$_{7/2}$ level of Yb$^{3+}$~\cite{Du_ecs}.
The Fermi energies were estimated to be 1.03 eV and 2.76 eV above VBMs for PBEsol and HSE06 respectively (see Table S2).
Then, the level of the optically excited state $^{2}$F$_{5/2}$ of Yb$^{3+}$ can be found at 1.26 eV (corresponding to 980 nm excitation energy) over this level, as denoted by yellow color in Fig.~\ref{fig2}.
When compared HSE06 calculation to the PBEsol, the transition levels are separated further from the $^{2}$F$_{5/2}$ level, indicating that only the hybrid functional can ensure the effective role of \ce{CaF2} in the UC/DC luminescence.
In the HSE06 calculation, the point defect Yb$_\text{Ca}$ and the pair defect Yb$_\text{i}$-F$_\text{i}$ are found to cause electron trap readily via non-radiative relaxation of absorbing one or two photons, due to their transition levels $\varepsilon$(1+/0) and $\varepsilon$(2+/1+) close to the $^{2}$F$_{5/2}$ level.
For the efficient energy transfer from \ce{CaF2} host to Yb$^{3+}$ ions, they should be suppressed.

Next, we calculated the formation energies of defects to examine their formation possibility.
%
%
%
The chemical potentials of species $\mu_{i}$ can be changeable according to the synthesis condition, and thus have the values between its lower (poor condition) and upper (rich condition) limits~\cite{Walle}.
We refer the rich conditions of Ca and F species to the fcc-metal bulk and gas states respectively, using the DFT total energies per atom.
In thermodynamic equilibrium growth conditions, the existence of \ce{CaF2} should satisfy $\mu_\ce{Ca}+2\mu_\ce{F}=\mu_\ce{CaF2}$, where $\mu_\ce{CaF2}$ is approximated to be the DFT total energy per formula unit of \ce{CaF2} unit cell, giving the constraint for the chemical potentials and their lower limits as $\mu^\text{poor}_\text{Ca}=\mu_\ce{CaF2}-\mu^\text{rich}_\text{F}$, $\mu^\text{poor}_\text{F}=\mu_\ce{CaF2}-\mu^\text{rich}_\text{Ca}$.
The relative differences between the lower and upper limits, $\Delta\mu_{\text{i}}=\mu^{\text{poor}}_{\text{i}}-\mu^{\text{rich}}_{\text{i}}$, were calculated to be $-$13.10 eV for Ca and $-$6.55 eV for F respectively.
For the extrinsic Yb dopant, besides the upper constraint from the metal bulk energy, another constraint exists as $\mu_\ce{Yb}+3\mu_\ce{F}\leq\mu_\ce{YbF3}$, which prevents the formation of secondary phase and sets up the upper limit of $\mu_\ce{Yb}$ according to the rich or poor condition of host elements.
Considering these two upper constraints, the difference between the chemical potentials for the F-rich and -poor conditions was obtained to be $-$14.04 eV.

%
\begin{figure}[!b]
\begin{tabular}{l}
\includegraphics[clip=true,scale=0.45]{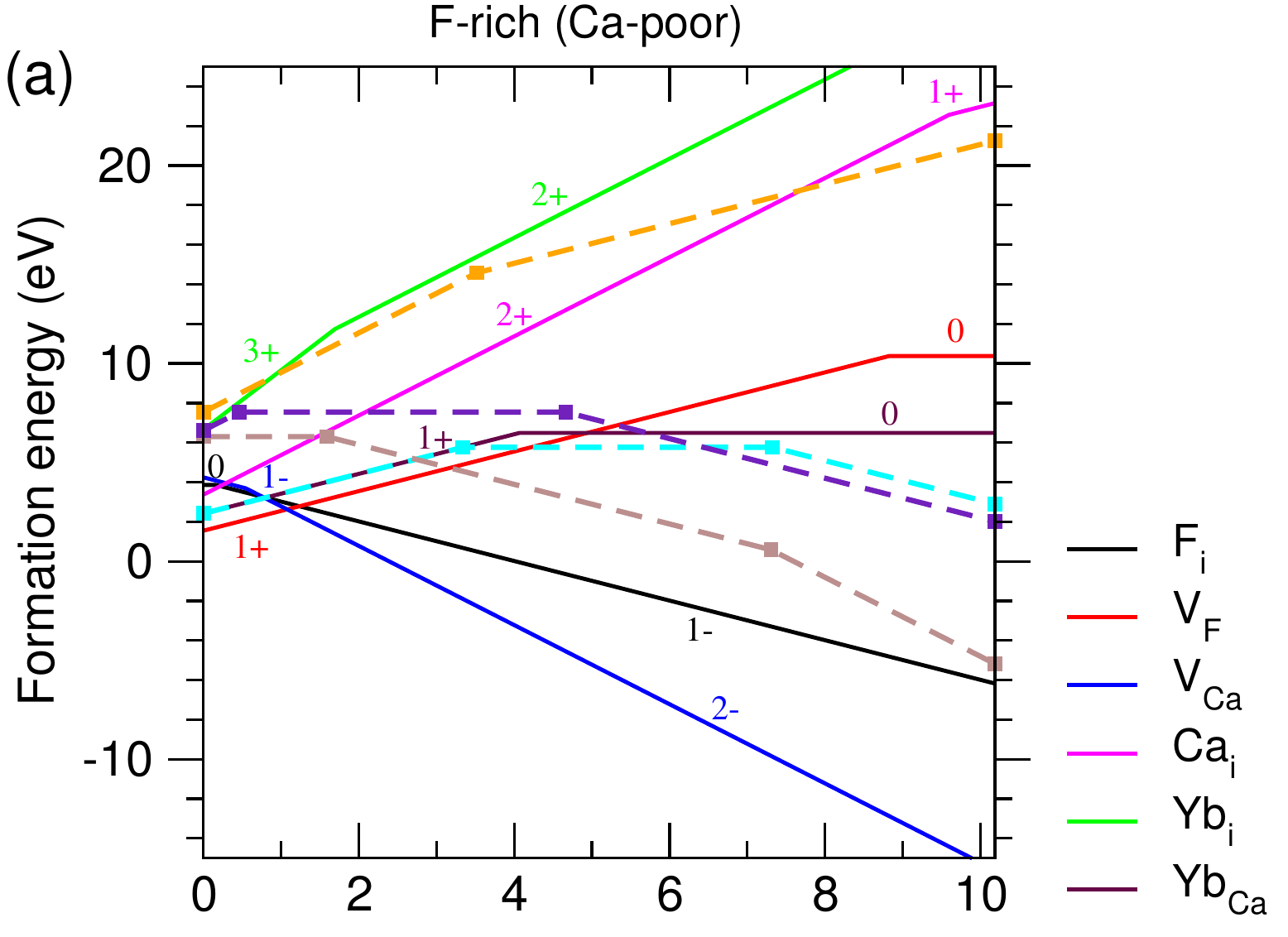} \\ \includegraphics[clip=true,scale=0.45]{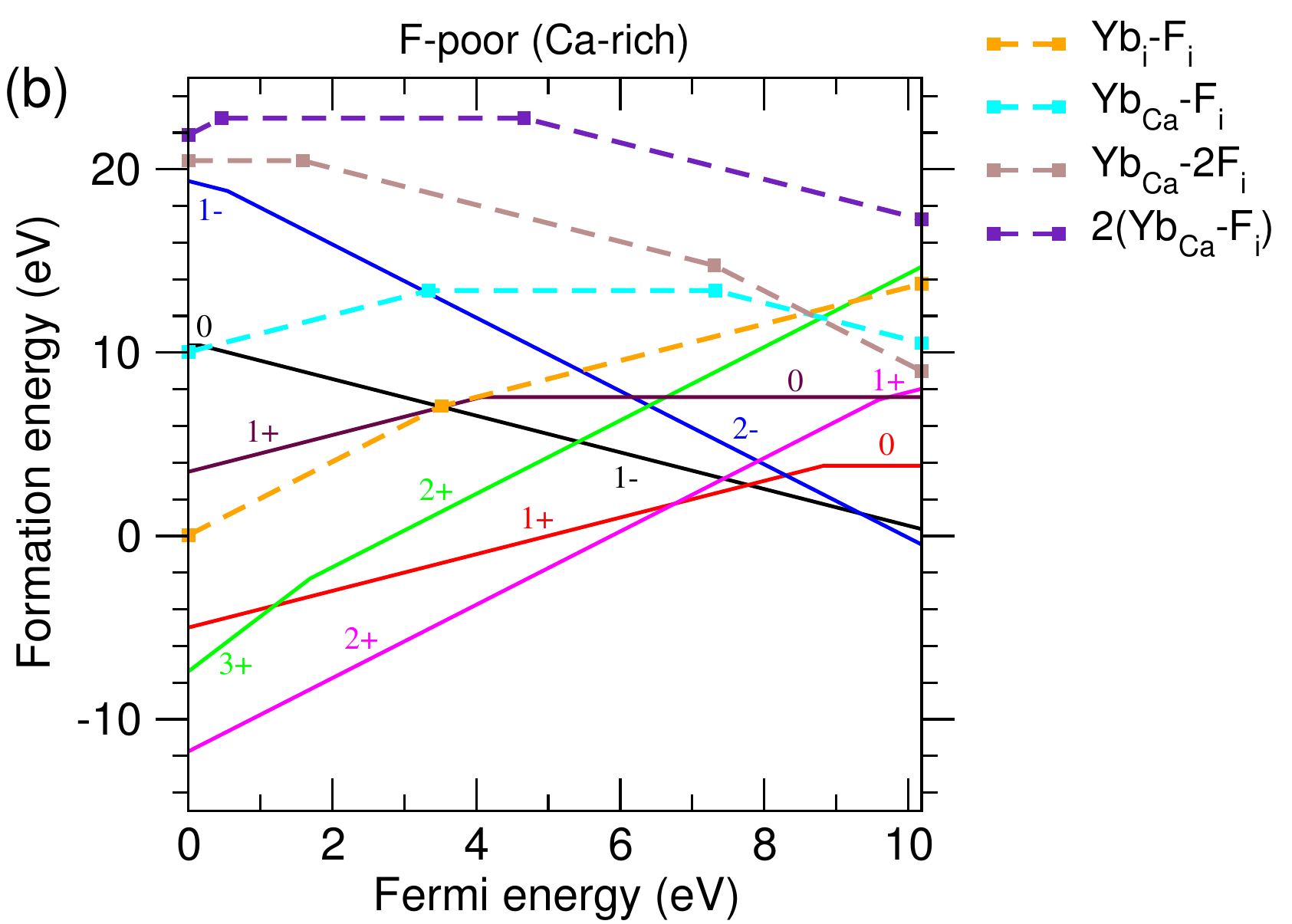}
\end{tabular}
\caption{\label{fig3}Defect formation energy diagrams under (a) F-rich (Ca-poor) and (b) F-poor (Ca-rich) conditions as a function of Fermi energy.}
\end{figure}
Figure~\ref{fig3} shows the formation energy diagrams of the relevant defects under the F-rich (Ca-poor) and F-poor (Ca-rich) conditions.
Under the F-rich condition, V$_\text{Ca}^{2-}$ is found to be the dominant defect with the lowest formation energy in the most range of $E_\text{F}$, while under the F-poor condition the point defect Ca$_\text{i}^{2+}$ is dominant.
Fortunately, these defects are not likely to create deep trap states.
Our major concern is the Yb-doped point defects.
Under the both conditions, Yb is in the possibly rich conditions as mentioned above, which, otherwise, can seem to be relatively poor in F-rich condition and relatively rich in F-poor condition.
Meanwhile, for the lower Yb concentrations, the formation energy diagrams of Yb-doped defects can be expected to move upwards.
As shown in Fig.~\ref{fig3}, the antisite Yb$_\text{Ca}$, which was found to generate deep trap states, has the formation energy bigger than 1 eV under any thermodynamically reasonable condition.
This is contradictory to the experimental results of high Yb-doping concentration~\cite{Zhou_jmcc} and so we conclude that Yb$_\text{Ca}$ defect can be feasibly formed only by the influence of the intrinsic defect F$_\text{i}^{1-}$.
On the other hand, the formation energy of interstitial Yb dopant (Yb$_\text{i}$) was found to be much higher on the whole span of $E_\text{F}$ under the F-rich condition but lower below $E_\text{F}=6.64$ eV under the F-poor condition than that of Yb$_\text{Ca}$.
Providing that Yb$_\text{Ca}$ is easily transformed to the pair defects like Yb$_\text{Ca}$-F$_\text{i}$, we can introduce the F-rich condition to suppress the formation of Yb$_\text{i}$ and the undesirable pair defect Yb$_\text{i}$-F$_\text{i}$.

\begin{figure}[!t]
\includegraphics[clip=true,scale=0.45]{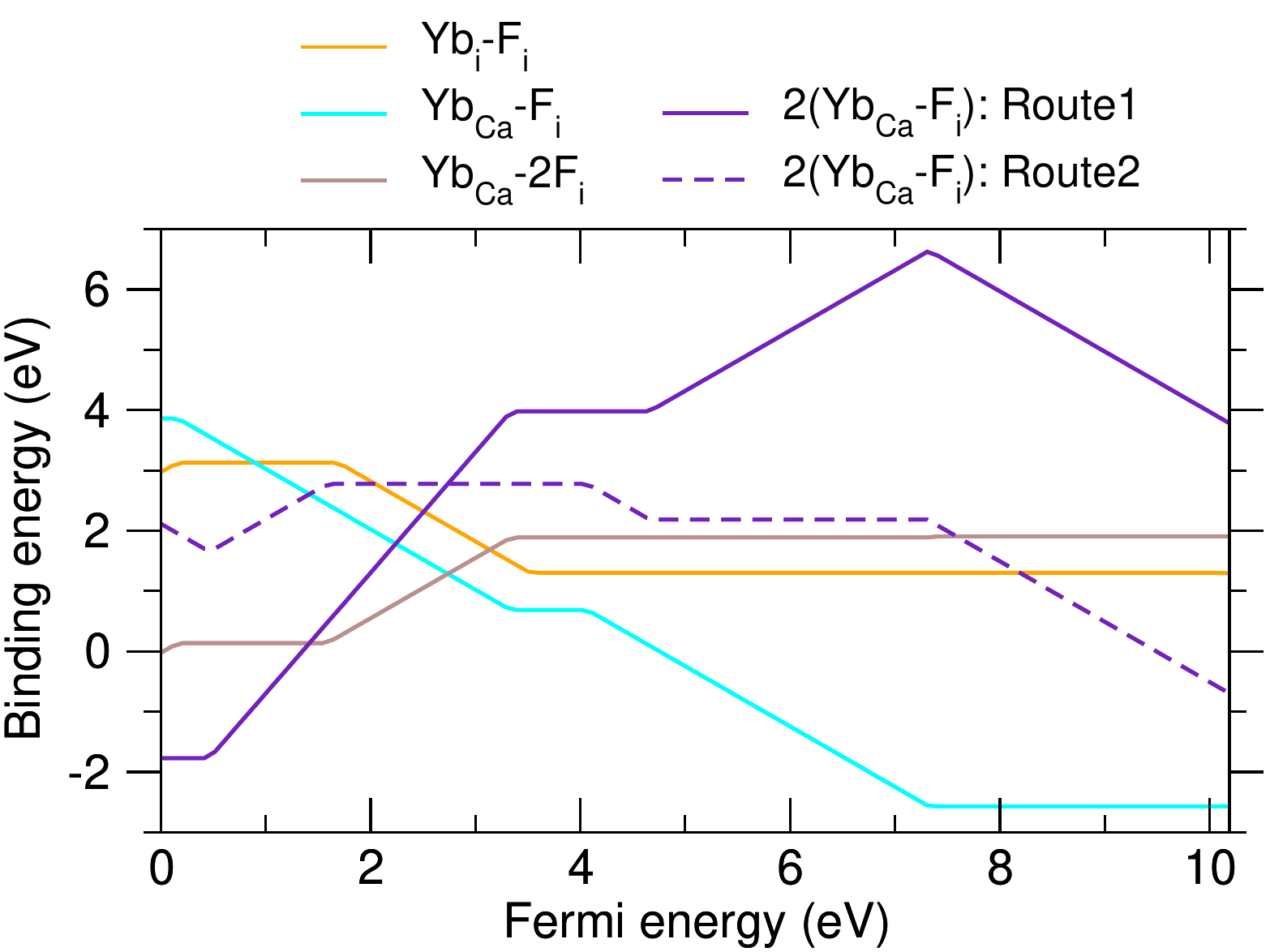}
\caption{\label{fig4}Binding energies of pair defects from their component point defects as a function of Fermi energy.}
\end{figure}

In order to check the probability of pair defect formation, we estimated their binding energies, defined as $E_\text{b}=H_\text{f}[A]+H_\text{f}[B]-H_\text{f}[AB]$~\cite{Walle}, where $H_\text{f}[A]$ is the formation energy of point defect $A$ and $H_\text{f}[AB]$ is that of pair defect.
For the complex defects composed of more than three components such as Yb$_\text{Ca}$-2F$_\text{i}$ and 2(Yb$_\text{Ca}$-F$_\text{i}$), the pair defect Yb$_\text{Ca}$-F$_\text{i}$ was adopted as a reactant, and for the latter case two possible reaction pathways were considered; Route1: Yb$_\text{Ca}$-F$_\text{i}$+Yb$_\text{Ca}$-F$_\text{i}$~$\rightarrow$~2(Yb$_\text{Ca}$-F$_\text{i}$) and Route2: Yb$_\text{Ca}$-2F$_\text{i}$+Yb$_\text{Ca}$~$\rightarrow$~2(Yb$_\text{Ca}$-F$_\text{i}$).
Figure~\ref{fig4} shows the calculated binding energies of the relevant pair defects.
Positive binding energy indicates the stable aggregation of defects and parallel parts of lines to the Fermi energy axis represent the charge equality of reactants and products.
Via the route of Yb$_\text{Ca}^{1+}$ + F$_\text{i}^{1-}$~$\rightarrow$~[Yb$_\text{Ca}$-F$_\text{i}$]$^{0}$ around  $E_\text{F}$~=~3.5~eV, Yb$_\text{Ca}$-F$_\text{i}$ has the positive binding energy of 0.68 eV, and for Yb$_\text{Ca}$-F$_\text{i}$, the route Yb$_\text{i}^{3+}$ + F$_\text{i}^{1-}$~$\rightarrow$~[Yb$_\text{i}$-F$_\text{i}$]$^{2+}$ yields the binding energy of 1.93 eV around $E_\text{F}=1$ eV.
Two routes for the formation of 2(Yb$_\text{Ca}$-F$_\text{i}$) from [Yb$_\text{Ca}$-F$_\text{i}$]$^{0}$ also verify the exothermic binding reaction as early suggested by Catlow's calculation~\cite{Catlow_prb} and elucidate the origins of experimentally observed Yb-clustering even at low concentration~\cite{Petit_prb, Lyberis_opt}.
Positive binding energies of Yb$_\text{i}$-2F$_\text{i}, $Yb$_\text{Ca}$-2F$_\text{i}$ and 2(Yb$_\text{Ca}$-F$_\text{i}$) indicate a facile aggregation of isolated Yb$_\text{Ca}$ with F$_\text{i}$ and guarantee the rare existence of this kind of trap defects.
Meanwhile, another quenching candidate Yb$_\text{i}$-F$_\text{i}$ can be favorably formed due to the positive binding energy through the whole $E_\text{F}$ range, requiring an induction of the F-rich condition for the high luminescence efficiency as mentioned above.

%

%

Finally, we turn our attention to the binding energy variation of complex defects in relation with the Fermi energy.
In Fig.~\ref{fig4}, it is noticeable that binding energy of Yb$_\text{Ca}$-F$_\text{i}$ decreases to a negative value of $-$2.57 eV as $E_\text{F}$ approaching to CBM.
Dimer binding energy in Route2 also shows a decreasing trend to the negative value near CBM, while that in Route1 is $-$1.77 eV near VBM.
Aggregation of F$_\text{i}$ and Yb$_\text{Ca}$-F$_\text{i}$ to Yb$_\text{Ca}$-2F$_\text{i}$ is not likely to occur around VBM because of its lower binding energy in the vicinity of zero.
In principle, this degradable behavior near the band edges is mainly originated in the different charge states of Yb and F dopants from their nominal values.
It is well accepted that the driving force of Yb-clustering is Coulomb interaction between Yb$_\text{Ca}^{1+}$ and F$_\text{i}^{1-}$, but stabilized neutral states of Yb$_\text{Ca}^{0}$ and F$_\text{i}^{0}$ impede the defect aggregation.
For this reason, we recommend extrinsic or intrinsic $p$-/$n$-type doping for controlling the Yb-clustering.
We can deduce that segregation product of $p$-type doping through the inverse process of Route1 is Yb$_\text{Ca}$-F$_\text{i}$, while $n$-type doping results in Yb$_\text{Ca}$-2F$_\text{i}$ following the inverse reaction of Route2.

In summary, we have investigated the defect physics of Yb-doped \ce{CaF2} to reveal thermodynamic transition levels and defect formation energies by using DFT calculations.
Multielectronic levels of Yb$^{3+}$ ion for the optical spectra ($^{2}$F$_{7/2}$ and $^{2}$F$_{5/2}$) were presumably determined in relation with the DFT electronic structure, and Yb$_\text{Ca}$ and Yb$_\text{i}$-F$_\text{i}$ were thought to be bulk energy quenchers with deep trap levels.
Under the F-rich growth condition, both trap defects can be suppressed due to the low concentration of Yb$_\text{i}$ and the dimerization of Yb$_\text{Ca}$.
Experimentally observed Yb-clustering effect was explained by the positive binding energies of defects, but $n$- or $p$-type doping provided negative effects on the aggregation.
Although further study on the binding mechanism between Yb and other Ln-dopants are needed, our study can provide a guidance for tuning of Ln-clustering in \ce{CaF2} for efficient UC/DC luminescence.

\section*{Acknowledgments}
This work was supported partially by the State Committee of Science and Technology, Democratic People's Republic of Korea.
The calculations have been performed on the HP Blade System C7000 (HP BL460c) that is owned and managed by the Faculty of Materials Science, Kim Il Sung University.

\section*{Appendix A. Supplementary data}
Supplementary data related to this article can be found at URL.

\section*{\label{note}Notes}
The authors declare no competing financial interest.

\bibliographystyle{elsarticle-num-names}
\bibliography{Reference}

\end{document}